\documentclass[10pt,a4paper]{article}
\usepackage{epsfig}
\usepackage{changes}
\columnwidth\textwidth

\begin{document}
\input{pstricks}

\pagestyle{plain}
\large
\vspace{50mm}
\begin{center}
\LARGE
{\bf Hadronic Final States in e$^{+}$e$^{-}$-Interactions and their Relevance to
Cosmic Rays} \\~\\
\end{center}
\vspace{1mm}
\Large
\begin{center}
{\bf Claus Grupen}

\enlargethispage{3cm}
\normalsize
{\it University of Siegen, Germany}
\end{center}
\vspace{11mm}
\setlength {\textwidth}{100mm}
\large
\begin{center}
{\bf ABSTRACT} \\
\begin{minipage}[t]{115mm}
\small
Elementary particle aspects using beams from accelerators and cosmic rays are
compared. An important feature in high energy physics is particle identification
in the final state. This technique is used to describe global features of
e$^{+}$e$^{-}$-annihilations such as multiplicities and multiplicity
distributions of charged hadrons and the relative fractions of pions, kaons and
protons. Also the characteristic properties of quark and gluon jets and scaling
violation in e$^{+}$e$^{-}$-interactions are discussed.\\~\\
\end{minipage}
\end{center}
\vspace{1mm}
\large
{\flushleft\bf INTRODUCTION} \\
\normalsize
~\\
Cosmic ray experiments investigating problems of high energy physics are always
fixed target experiments. This implies the disadvantage that only a small
fraction of the energy of the incident particle can be used for the creation of
new particles while the major part is used to boost the interaction products
into the forward direction.

For a target of mass $m$ and an energy of the incident particle $E_{\rm lab}$
the center-of-mass energy squared $s$ is calculated to be

\begin{equation}
s=2mE_{\rm lab},
\end{equation}
~\\
as long as $E_{\rm lab}\gg mc^{2}$. This leads to a center-of-mass energy of
$1\,{\rm TeV}$ for a cosmic ray particle of $5\cdot 10^{14}\,{\rm eV}$ hitting a
proton target.

The best use of the energy of the particle at accelerators is made with the
technique of storage rings. Two particles of $500\,{\rm GeV}$ each achieve the
same center-of-mass energy of $1\,{\rm TeV}$ in a head-on collision. Here the
full energy of the particles can be used for new particle creation.

\begin{figure}[hb]
\begin{center}
{\epsfig{figure=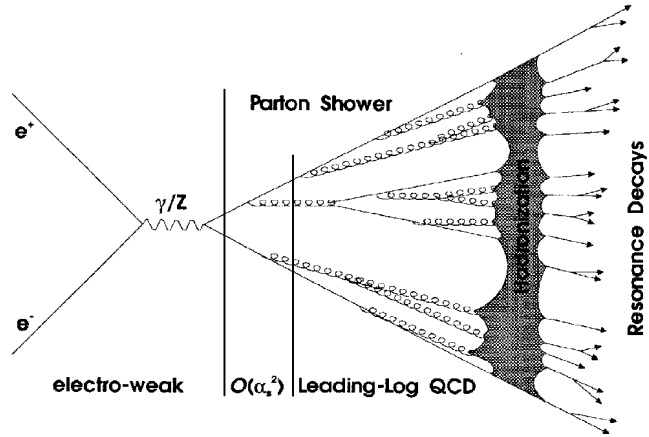,height=5.9cm}}
\end{center}
\caption[]{\tt Fundamental processes of hadron production in e$^{+}$e$^{-}$
annihi-\linebreak
lation~\cite{bethge,schmelling}.}
\label{fig1}
\end{figure}

The particle accelerators in operation and those in construction will cover
equivalent laboratory energies up to $10^{17}\,{\rm eV}$. The energy domain
beyond $10^{17}\,{\rm eV}$ will\linebreak
~\\
~\\
~\\
~\\
{\large
\begin{center}
\noindent
Conf. Proc. XV Cracow Summer School on Cosmology, July 15-19, 1996, Lodz,
Poland.
\end{center}}
\clearpage
\noindent
be reserved for high energy physics with cosmic
rays. The problem there, however, is that the rates are extremely low. But
elementary particle physics is only one aspect of cosmic rays, the other
possibly more important one is astronomy and astrophysics.
\\
\\
\large
{\bf\boldmath EXPERIMENTAL TECHNIQUES IN e$^{+}$e$^{-}$ COLLIDER-\linebreak
EXPERIMENTS}
\newline
\normalsize
~\\
The fundamental process of hadron production in e$^{+}$e$^{-}$-collisions can be
subdivided into five steps:
\begin{description}
\item[(i)]
The electron and positron annihilate into a virtual gauge boson, $\gamma$ or Z;
this process is entirely described by the standard model of electroweak
interactions (see Fig.~\ref{fig1},~\cite{bethge,schmelling}).
\item[(ii)]
The photon or Z produce a quark-antiquark pair. This transition is also
understood in the framework of the standard model.
\item[(iii)]
The quarks initiate a quark-gluon cascade. As long as the virtuality of the
partons is larger than a certain cut-off value (typically $1\,{\rm GeV}$) this
parton shower can be calculated in perturbative quantum chromodynamics (QCD).
\item[(iv)]
The transition of partons to primary hadrons cannot be calculated in QCD but
rather requires phenomenological models of the hadronization process.
\item[(v)]
Finally the primary hadrons or their decay products
($\rho^{0}\rightarrow\pi^{+}\pi^{-}$; $\Lambda\rightarrow p\pi^{-}$;
$\eta\rightarrow\gamma\gamma$; $\ldots$)
can be recorded in the detector~\cite{webber}.
\end{description}

It is desirable to identify all particles in the hadronic final state. Several
techniques can be used to achieve this~\cite{grupen}:

\begin{description}
\item[(i)]
The ionization loss ${\rm d}E/{\rm d}x$ provides information on the charge $Z$
of the particle (usually $Z=1$ in e$^{+}$e$^{-}$ collisions) and its velocity
$\beta$.
\item[(ii)]
Ring-imaging Cherenkov counters (RICH) equally determine $Z$ and $\beta$.
\item[(iii)]
Electromagnetic and hadron calorimeters distinguish between e$^{+}$, e$^{-}$,
$\gamma$ on the one hand and $\pi$, K, p, n on the other hand on the basis of
the characteristically different shower profiles of the particles.
\item[(iv)]
Muons are identified by their penetration through e.g. the hadron calorimeter.
\item[(v)]
Particles which decay after a very short path length are identified by the
reconstruction of their invariant mass from their decay products.
\item[(vi)]
Finally an accurate momentum measurement in a high resolution spectrometer
$p=\gamma m_{0}\beta c$ fixes the particle mass, if its velocity is known.
\end{description}

\begin{figure}[hb]
\begin{center}
{\epsfig{figure=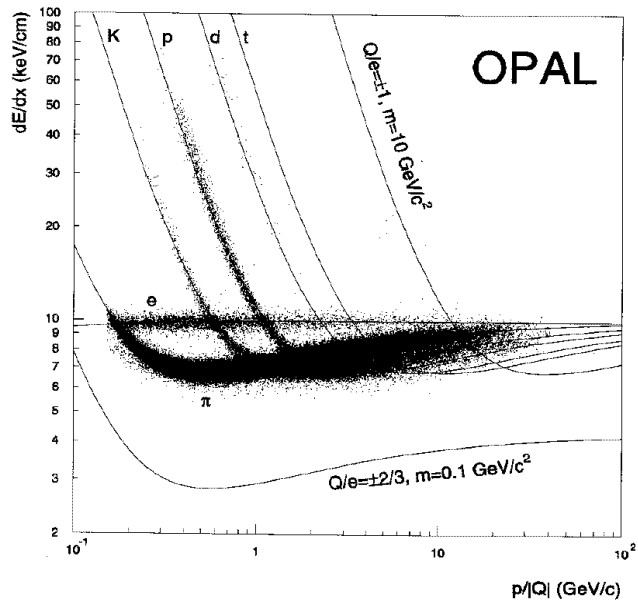,height=8.5cm,width=9.0cm}}
\caption[]{\tt Energy\hspace{0.1mm} loss\hspace{0.1mm} curves\hspace{0.1mm}
for\hspace{0.1mm} electrons,\hspace{0.2mm} pions,\hspace{0.2mm}
kaons,\hspace{0.2mm} protons\hspace{0.1mm} and

deuterons in the OPAL central tracking chamber~\cite{opal2}. The dia\-gram al-

so\hspace{0.2mm} shows\hspace{0.2mm} that\hspace{0.2mm} exotic\hspace{0.2mm}
particles\hspace{0.2mm} are\hspace{0.2mm} not\hspace{0.2mm}
produced\hspace{0.2mm} in\hspace{0.2mm} e$^{+}$e$^{-}$\hspace{0.2mm}
interacti-

ons.}
\label{fig2}
\end{center}
\end{figure}

Fig.~\ref{fig2} shows the energy loss curves for electrons, pions, kaons,
protons, deuterons and tritons in the OPAL tracking chamber~\cite{opal1,opal2}.
Below momenta of $1\,{\rm GeV}/{c}$ a clear separation between the different
particle species is possible. Also in the relativistic rise region a certain
particle identification power is obtained. This can be substantially improved
upon by making use of additional information from calorimeters and\,/\,or
Cherenkov counters.
~\\
\\
\large
{\bf MULTIPLICITIES AND MUL\-TI\-PLI\-CITY DIS\-TRI\-BU\-TIONS} \newline
\normalsize
~\\
At first glance the determination of the charged multiplicity in hadronic final
states seems to be easy. However, the loss of particles at low momentum and
imperfections of track reconstruction in collimated jets requires an unfolding
procedure to arrive at the true multiplicity from the observed
one~\cite{aleph1}.

\begin{figure}[hb]
\begin{center}
{\epsfig{figure=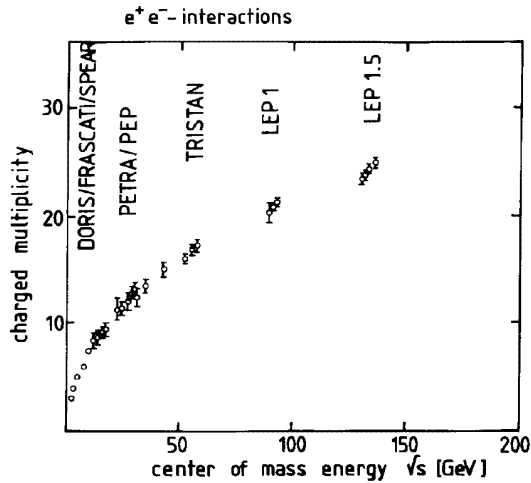,width=7.2cm,height=6.65cm}}
\end{center}
\caption[]{\tt Charged multiplicities in e$^{+}$e$^{-}$ interactions from
various e$^{+}$e$^{-}$ collider experiments (after \cite{schmelling}).}
\label{fig5}
\end{figure}

Fig.~\ref{fig5} shows the charged particle multiplicities observed in
e$^{+}$e$^{-}$-interactions at different center-of-mass energies
(after~\cite{schmelling}). The data can be compared to various
parametrizations~\cite{opal3}. The formulae given in the following equations
(\ref{eq3})\,-\,(\ref{eq8}) are based on fits to data obtained in the energy
range from $\sqrt s=2\,{\rm GeV}$ to $\sqrt s=91\,{\rm GeV}$. Recently obtained
results at $\sqrt s=133\,{\rm GeV}$ can be used to judge on the quality of the
parametrization~\cite{opal4,aleph2,delphi,l3}.

A simple power law

\begin{equation}
\left<n_{\rm ch}\right>=a\cdot s^{b}
\label{eq3}
\end{equation}
~\\

{\noindent
motivated by phase space arguments does not fit the new data well ($a=2.22$;
$b=0.252$). Fermi's idea applied to discrete distributions}

\begin{equation}
\left<n_{\rm ch}\right>=\beta\cdot s^{\alpha}-1
\end{equation}
~\\
describes the data rather well ($\beta=2.979$; $\alpha=0.222$), and so does the
frequently used empirical relation

\begin{equation}
\left<n_{\rm ch}\right>=a+b\cdot{\rm ln}s+c\,{\rm ln}^{2}s
\end{equation}
~\\
($a=3.297$; $b=-0.394$; $c=0.263$). The QCD inspired parametrization based on
leading-log approximation

\begin{equation}
\left<n_{\rm ch}\right>=a+b\cdot{\rm e}^{c\cdot\sqrt{{\rm
ln}\left(s/Q_{0}^{2}\right)}}
\end{equation}
~\\
with $a=2.418$; $b=0.113$; $c=1.712$ and $Q_{0}=1\,{\rm GeV}$ cannot really
account for the $133\,{\rm GeV}$ data, but the inclusion of next-to-leading-log
expressions in QCD provides a good description of the high energy data

\begin{equation}
\left<n_{\rm ch}\right>=a\cdot\alpha_{s}\,^{b}\cdot{\rm e}^{c/\sqrt{\alpha_{s}}}
\end{equation}
~\\
with $a=0.065$; $b=0.49$; $c=2.27$ and
~\\
\begin{equation}
\alpha_{s}=\frac{4\pi}{\beta_{0}\,{\rm ln}s/{\Lambda^{2}}}-\frac{4\pi\beta_{1}\,
{\rm ln}\,{\rm ln}s/{\Lambda^{2}}}{\beta_{0}\,^{3}\,{\rm ln}^{2}s/{\Lambda^{2}}}
\label{eq8}
\end{equation}
~\\
\noindent
with $\beta_{0}=7.67$; $\beta_{1}=38.67$ and $\Lambda=136\,{\rm MeV}$.
$\alpha_{s}$ is the running coupling constant of strong interactions.

For the same center-of-mass energy the charged particle multiplicity in
proton-proton collisions is smaller compared to e$^{+}$e$^{-}$-collisions. This
relates to the fact that in e$^{+}$e$^{-}$-interactions the full center-of-mass
energy is available for particle production while in pp-collisions one deals
with quark-quark scattering, and the quark-quark system carries only a fraction
of the proton-proton center-of-mass energy~\cite{pdg}.

\begin{figure}[ht]
\begin{center}
{\epsfig{figure=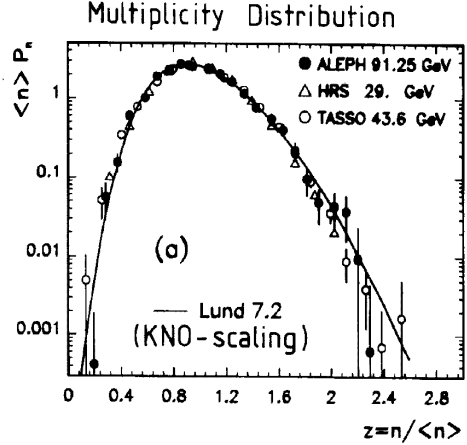,height=6.7cm,width=7.0cm}}
\end{center}
\caption[]{\tt Unfolded charged multiplicity distribution in KNO form com-

paring data from ALEPH, HRS and TASSO~\cite{aleph1}}
\label{fig6}
\end{figure}

The multiplicity distribution (Fig.~\ref{fig6}) exhibits KNO-scaling when
plotted in the normalized variables $\Psi(z)=\left<n_{\rm ch}\right>\cdot
P(n_{\rm ch})$ and $z=n_{\rm ch}/{\left<n_{\rm ch}\right>}$, where $P(n_{\rm
ch})$ is the probability to observe $n_{\rm ch}$ hadrons for an average of
$\left<n_{\rm ch}\right>$~\cite{aleph1}.
For a center-of-mass energy of $\sqrt s=91\,{\rm GeV}$ the multiplicity
distribution is best described by a log-normal distribution, but also the
negative binomial distribution fits the data quite well.

At low momenta $(\leq 1\,{\rm GeV}/c)$ the final state is dominated by pions
$(\geq 85\,\%)$. The particle composition changes significantly for higher
momenta. Around $p=10\,{\rm GeV}/c$ there are about $60\,\%$ pions, $30\,\%$
kaons and $10\,\%$ protons.\linebreak
\clearpage
\noindent
The high statistics also allows to identify the
production of strange and charmed baryons, such as $\Sigma$, $\Xi$ and $\Omega$.
For example, the $\Omega^{-}$ multiplicity is of the order of $0.001$ per event
at $\sqrt s=91\,{\rm GeV}$.
\\
~\\
\large
{\bf PROPERTIES OF QUARK AND GLUON JETS}
\newline
\normalsize
~\\
In three-jet events, where one of the primary quarks radiates an energetic gluon
it is possible, to distinguish the gluon jet from the quark jets. If one selects
${\rm q}\bar{\rm q}{\rm g}$ events where the quarks are either charm or bottom
quarks one may find in these quark jets evidence for long-lived hadrons
(B-mesons or D-mesons). Experimentally one looks for decay vertices of B or
D-mesons displaced from the primary vertex. In the gluon jet the production of
B or D-mesons is rather unlikely because the charm and bottom quarks are too
heavy. Consequently in three-jet events with evidence for long-lived hadrons in
two jets the third jet must have been initiated by a gluon.

The sample of tagged three-jet events can now be used to search for differences
between quark and gluon jets. It turns out that gluon jets are wider and contain
also more particles than quark jets. The increase in particle multiplicity in
gluon jets is concentrated at low hadron momenta causing the fragmentation
function of gluons to be softer than the quark fragmentation
function (e.g.~\cite{opal5}).
\\
~\\
\large
{\bf SCALING VIOLATIONS IN e$^{+}$e$^{-}$-INTERACTIONS}
\newline
\normalsize
~\\
Feynman-scaling states that in the quark-parton model the normalized inclusive
cross-section

\begin{equation}
\frac{1}{\sigma_{\rm tot}}\,\frac{{\rm d}\sigma}{{\rm
d}x}\hspace{7mm}{\rm with}\hspace{7mm}x=\frac{E_{\rm hadron}}{E_{\rm
beam}}=\frac{2\,E_{\rm hadron}}{\sqrt s}
\end{equation}
~\\
does not depend on the center-of-mass energy. Feynman-scaling is known to be
violated, and the evidence for that comes from accelerator and cosmic ray
experiments. The reason for this scaling violation originates from gluon
radiation which leads to a dependence of $\frac{1}{\sigma_{\rm tot}}\,\frac{{\rm
d}\sigma}{{\rm d}x}$ on the center-of-mass energy.

These scaling violations come about because with increasing $\sqrt s$ more phase
space for gluon radiation and thus final state particle production becomes
available leading to a softer $x$-distribution at higher center-of-mass
energies.

\begin{figure}[hbt]
\begin{center}
{\epsfig{figure=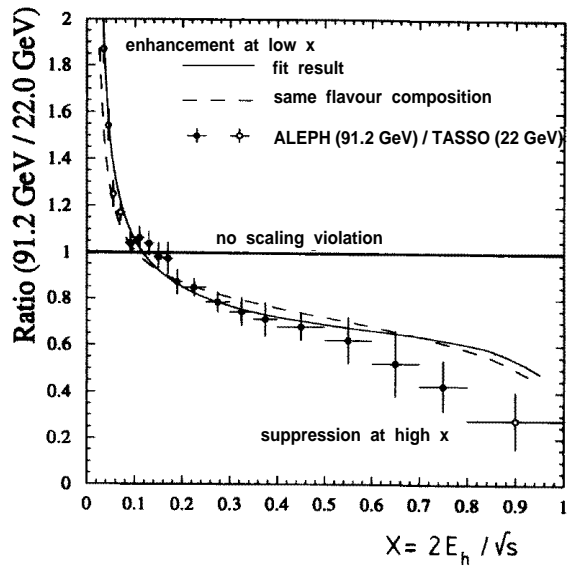,height=8.0cm,width=8.5cm}}
\end{center}
\caption[]{\tt Ratio of inclusive cross-sections at
$\sqrt{\tt s}={\tt 91.2}$\,GeV and $\sqrt{\tt s}={\tt 22}\,{\tt
GeV}$~\cite{aleph4}.}
\label{fig8}
\end{figure}

The inclusive cross-section for

\begin{equation}
{\rm e}^{+}{\rm e}^{-}\rightarrow{\rm hadron\,+\,anything}
\end{equation}
~\\
has been measured for center-of-mass energies between $22\,{\rm GeV}$ and
$91\,{\rm GeV}$~\cite{schmelling,aleph4}. The fragmentation functions get softer
at higher values of $\sqrt s$. This is clearly seen in Fig.~\ref{fig8} where the
ratio of the inclusive cross-sections at $91\,{\rm GeV}$ and $22\,{\rm GeV}$ is
shown~\cite{aleph4}. Higher center-of-mass energies lead to an enhancement at
low momenta and a depletion at high hadron momenta.
~\\
\clearpage
\large
{\noindent\bf CONCLUSIONS}
\newline
\normalsize
~\\
Colliders can reach equivalent laboratory energies of $E_{\rm lab}=10^{17}\,{\rm
eV}$. The domain beyond this energy is reserved for cosmic rays.

Hadron production in e$^{+}$e$^{-}$ experiments is well described by the
standard model of electroweak interactions, quantum chromodynamics and
phenomenological models for the hadronization of quarks and gluons. It is highly
desirable to also understand the transition of partons to hadrons in the
framework of QCD.
\\
~\\
\large
{\bf ACKNOWLEDGEMENTS}
\newline
\normalsize
~\\
I am grateful for the hospitality and support provided by the summer school
organizers. My special thanks go to Janusz Kempa, Jerzy Wdowczyk and Wieslaw
Tkaczyk. I thank also Detlev Maier and Volker Schreiber for their help in
preparing the written version of my talk.

\end{document}